\documentclass{mem}
\usepackage{natbib}\usepackage{txfonts}\usepackage{balance}
\usepackage{graphicx}
\usepackage{txfonts}
\usepackage[a4paper]{hyperref}
\idline{XXXX}{1}
\begin{document}
\def\teff{$T\rm_{eff }$}
\def\kms{$\mathrm {km s}^{-1}$}

\title{
Abundance structure of the Galactic disk
}

   \subtitle{}

\author{
S. \,Feltzing
%\inst{1} 
          }

  \offprints{S. Feltzing}

\institute{
Lund Observatory, Box 43, SE-221 00 Lund, Sweden
\email{sofia@astro.lu.se}
}

\authorrunning{Feltzing}

\titlerunning{Abundance structure of the Galactic disk}

\abstract{The current knowledge of the abundance structure in the
  Milky Way is reviewed. Special emphasis is put on recent results for
  stars with kinematics typical of the thin and the thick disks,
  respectively, and how these results can, apart from studying the
  Milky Way also give information about the origin of the
  elements. Two on-going studies are high-lighted. One that
  presents new data for stars with kinematics typical of the Hercules
  stream and one for a small number of dwarf stars high above the galactic plane. 
These indicate that the thick disk is a homogeneous structure and that the 
Hercules stream perhaps is part of the metal-rich thick disk. Furthermore, new
abundance determinations for Mn for stars with kinematics typical of the thin and the thick disk are presented. From these results the suggestion is that the observed trends for Mn in the Milky Way disks can be explained by metallicity dependent SN\,II yields. 
The impacts of surveys on the studies of the elemental
  abundance trends in the Galaxy are discussed. It is argued that when
  it comes to furthering our understanding of the abundance structure
  in the Galactic disk future and ongoing surveys' major impact will
  be in the form of catalogues to select targets from for high resolution
spectroscopic follow-up. For a correct
  interpretation of the results from these follow-up studies it is
  important that the surveys have well understood completeness
  characteristics.    }
\maketitle{}

\section{Introduction}

The Galaxy is a large, complex entity and our current observational
resources for high-resolution spectroscopy limit us to study mainly
the stars that are closest to us. In the solar neighbourhood several
different stellar populations overlap. These populations show
different kinematic properties and it is therefore possible to, at
least statistically, disentangle them (\citealt{fbh}).

A combination of kinematic information and abundance results based on
high resolution spectroscopy has proved to be a very valuable
instrument to disentangle the histories and properties of these
various stellar populations. \citet{ed03} provided one of the first
such studies, subsequent investigations have shown the disk system to
be very complex indeed, see e.g. \citet{mashonkina03},
\citet{fuhrmann04}, \citet{bensby04}, and \citet{reddy06}. The
important thing about all of these studies is that they adopt a
differential methodology such that stars with different kinematic
signatures are analysed and studied with exactly the same methods in
the abundance analysis. This differential approach has been able to
reveal intriguing differences between kinematically distinct groups,
such as the thin and the thick disk (e.g. \citealt{mashonkina03},
\citealt{fuhrmann04}, and \citealt{bensby04}).

Here I will concentrate on what can be learned from studies of,
mainly, dwarf stars, however, we should keep in mind that stars at
other evolutionary stages also can be successfully used to study the
anatomy of the Galaxy. My reasons for focusing on the dwarf stars, with
spectral types close to that of the sun, is that their spectra are,
reasonably, easy to analyse and, perhaps most importantly, they are
the type of stars that we currently have the most extensive data sets
for.

\section{Abundance structure of the Galactic disk}
\label{examples}

\subsection{$\alpha$-elements} 

\citet{fuhrmann98} provided what is probably the first study that
directly compared, in a differential manner, the elemental abundance
trends for stars with kinematics typical of the thin and the thick
disk, respectively.  \citet{fuhrmann04} provides an update and
extension of the previous work. In summary what he found is that, at a
given [Fe/H], stars with kinematics typical of the thick disk are
enhanced in [Mg/Fe] as compared to stars that have kinematics typical
of the thin disk. This has subsequently been shown to be true for all
the other so called $\alpha$-elements, including oxygen.

\begin{figure}[]
\resizebox{\hsize}{!}{\includegraphics[]{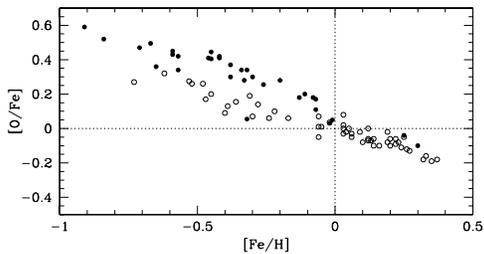}}
\caption{ \footnotesize [O/Fe] vs. [Fe/H] for two samples of
  stars. $\bullet$ denote stars that have kinematics typical of the
  thick disk whilst $\circ$ are stars with kinematics typical of the
  thin disk. Based on data from \citet{bensby04} and \citet{bensby05}.
}
\label{o}
\end{figure}

Oxygen is a particularly interesting element from the point of view of
chemical evolution in that it is essentially made exclusively in
SN\,II. Many elements have a combination of sources, e.g. iron is made
in both SN\,II and Ia. These types of supernovae operate on distinctly
different time scales. SN\,II originate in massive stars and have
progenitor lifetimes from tens to one hundred million years. In
contrast SN\,Ia are not yet fully understood apart from that they
arise from binary systems (see e.g. \citealt{livio01}). The lifetime
of an object that becomes a SN\,Ia therefore varies between
different models. For double degenerate system (i.e. made up of two
white dwarfs) the lifetime might be longer than the age of the
Universe. However, various recent studies have indicated the need for
more than one type of progenitor system. For example, in order to
explain the supernova rates in nearby galaxies \citet{mannucci}
suggest that both systems with a very short lifetime as well as systems
with long lifetimes are needed. The effect of such combinations of
SN\,Ia are not yet fully explored in the context of
Galactic chemical evolution (however, see e.g. \citealt{scannapieco}
for a first example of such modelling).

By comparing oxygen with an element such as iron it is thus possible
to deduce at what ``time'' the contribution from SN\,Ia sets in in our
Galaxy. With ``time'' I here mean some measure of time. It could
either be the age of the actual stars for which we see the signature
of SN\,Ia or it could e.g. be the iron content of such
stars. Figure\,\ref{o} shows such a plot for two kinematically
selected stellar samples. In brief the two samples represent stars
that have kinematics typical of either the thin or the thick disk. A
full description of the method used to select the stars is given in
\citet{bensby03} and an extended discussion of the effect of the local
normalisation may be found in \cite{bensby05}.

We see that the thick disk trend for [O/Fe] is flat at lower
metallicities but at around --0.5 dex it turns downwards. A
``knee'' can be seen. We see no such abrupt change in the trend for
the thin disk stars, but rather a gentle decline. These trends have
been interpreted such that the thick disk first experiences a strong
star formation period where iron is built up, and as only SN\,II are
operational the [O/Fe] ratio remains constant. At a given time, which
depends on the time-scales involved in the SN\,Ia production, the iron
production is increased through the contribution from SN\,Ia. As
SN\,Ia do not make oxygen this results in a downward trend. The story for
the thin disk is different -- there the star formation rate is lower,
and hence the contributions from SN\,II and SN\,Ia mix and we see no
sharp features in the abundance trends that (mainly) trace these two
contributions to the chemical enrichment.

\begin{figure}[]
\resizebox{\hsize}{!}{\includegraphics[]{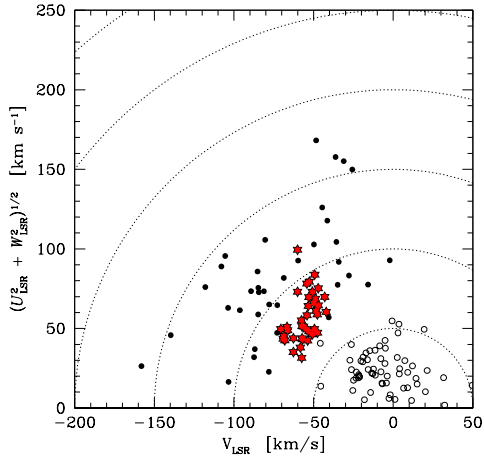}}
\caption{ \footnotesize Toomre diagram showing how the selected stars
  in the Hercules stream ($\bigstar$ grey/red) compares to the stars
  with kinematics typical of the thin ($\circ$) and thick disks
  ($\bullet$), respectively.  }
\label{hertoomre}
\end{figure}

\begin{figure}[]
\resizebox{\hsize}{!}{\includegraphics[]{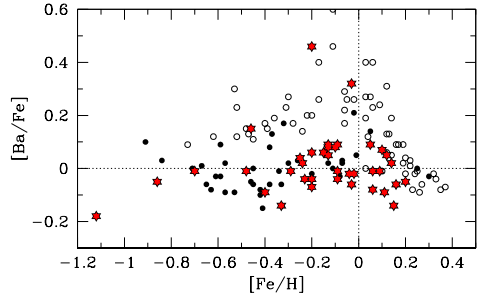}}
\caption{ \footnotesize [Ba/Fe] vs. [Fe/H] for stars in the Hercules
  stream. These stars are compared to local, kinematically selected,
  thin and thick disk stars \citep{bensby05}.  $\circ$ shows stars
  with kinematics typical for the thin disk, $\bullet$ stars with
  kinematics typical of the thick disk (data from
  \citealt{bensby05}). $\bigstar$ (grey/red) indicate the stars
  belonging to the Hercules stream. }
\label{her}
\end{figure}

\begin{figure}[]
\resizebox{\hsize}{!}{\includegraphics[]{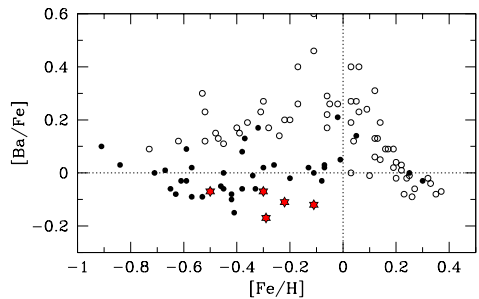}}
\caption{ \footnotesize Abundance plot for Ba for ``in situ'' dwarf
  stars and local thick and thin disk stars. $\circ$ shows stars with
  kinematics typical for the thin disk, $\bullet$ stars with
  kinematics typical of the thick disk (data from
  \citealt{bensby05}). $\bigstar$ (grey/red) indicate the five dwarf
  stars in the SGP field of \citet{gilmore95} that we have obtained
  high-resolution spectra for.  }
\label{insitufig}
\end{figure}

\subsection{Hercules stream}

It has long been recognised that the disk(s) contains several streams
or moving groups and Olin Eggen in particular explored these streams
at an early stage. The advent of Hipparcos and its superior data sets
provided a new incentive to revisit these streams again (e.g.
\citealt{fh00}).  The Hercules stream provides one example of such an
entity. The Hercules stream was recently identified as a dynamical
structure by \citet{famaey05} and \citet{navarro00}.

From investigations using the \citet{nordstrom04} catalogue we find
that the stars in the Hercules stream appears to have a mean metallicity
compatible with that of the thin disk. 

However, in a recent study of the elemental abundances in Hercules
stream stars \citet{bensbyher} find that the stars show chemical
signatures typical of the thick disk, see Fig.\,\ref{her}. This raises
questions about the origin of the stream -- do the stars come from the
Bulge, from the inner thick disk, or are they (partly?) part of a more
heated part of the thin disk? The first interpretation is less likely
as we know that the stars in the Galactic bulge are more enhanced in
the $\alpha$-elements than the Hercules stars are at [Fe/H]=0
\citep{fulbright}.

\subsection{Dwarf stars at 1-2 kpc}
\label{insitu}

Very few abundance studies have targeted dwarf stars well above the
Galactic plane. \citet{gilmore95} did an extensive investigation of
the metallicity distribution function at 1.5 and 2 kpc above the plane
using dwarf stars as tracers of the old stellar population.  They used
high-resolution, low S/N spectra to derive the stellar parameters and
found that there is no change in the metallicity distribution function
between these two samples. In a recent study we have selected a small
number of these faint dwarf stars ($V \sim 15-17$) for high-resolution
spectroscopy with UVES on VLT. The first results imply that the stars
show the distinct signatures of the thick disk, as observed in
kinematically selected sample in e.g. \citet{bensby05}. 
An example of our results are shown in Fig.\,\ref{insitufig} (Feltzing 
et al. in prep.).

This is not in itself surprising as the thick disk is the dominant
stellar population above $\sim 1$ kpc. However, this gives a further
indication that the thick disk consists of a homogeneous stellar
population pointing to a common origin for all its stars. Furthermore,
this gives additional constraints on any model of Galaxy formation and
evolution that sets out to explain the formation of (the ubiquitous?)
thick disks in spiral galaxies like our own \citep{yoachim}.

\section{Origin of elements}

Here I give two examples of how studies of kinematically distinct
stellar populations can help us to understand the nucleosynthetic
origin of the elements. The two examples are for carbon and manganese
and are taken from \citet{bf06} and \citet{feltzing06}.

\subsection{Carbon}

\begin{figure}[]
\resizebox{\hsize}{!}{\includegraphics[bb=18 144 592 485, clip=true]{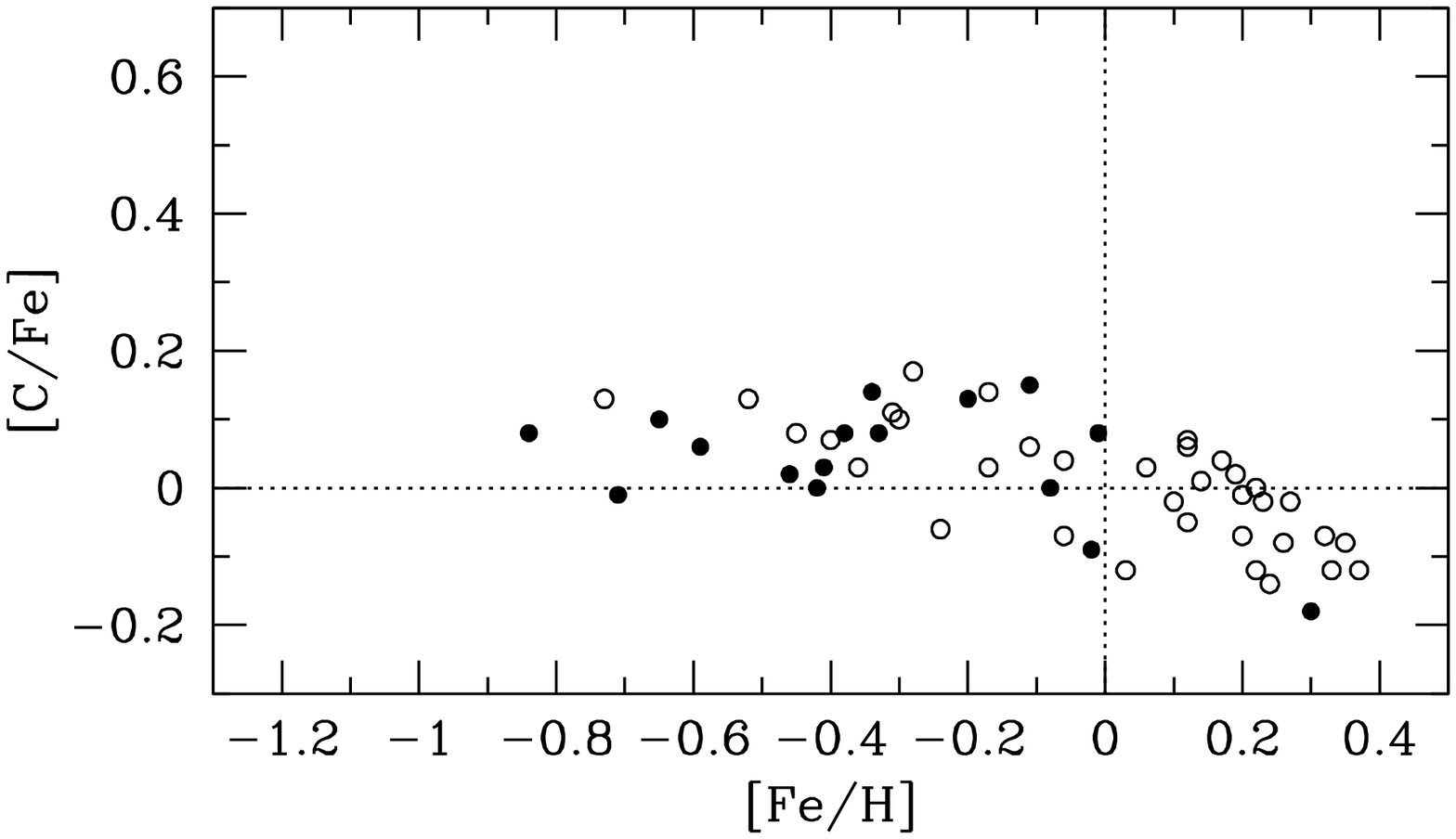}}
\caption{ \footnotesize [C/Fe] vs. [Fe/H] for stars in the thin ($\circ$) 
and thick ($\bullet$) disks, respectively.
 Carbon abundances are from \citet{bf06} and iron
  abundances have been taken from \citet{bensby03} and
  \citet{bensby05}.}
\label{c}
\end{figure}

Figure \ref{c} shows the abundance trends for carbon derived for 51
stars with typical thin or thick disk kinematics (\citealt{bf06}).
\citet{bf06} used a forbidden carbon line for the abundance analysis
and hence the results are robust against departure from local
thermodynamic equilibrium.  For [C/Fe] as a function of [Fe/H] the
thin and the thick disk trends are fully merged. At super-solar
metallicities there is an indication that the trend for the thin disk
starts to drop.

Comparing with Fig.\,\ref{o} we note that there is no change in the
thick disk carbon trend at the [Fe/H] where the downturn in [O/Fe]
happens.  Although we know that carbon is not made in the same sites
as oxygen it does, however, show that the source(s) that produce
carbon are operating on the same time-scales as those providing the
iron, i.e. the SNIa.

This, in principle, means that if we know where carbon is made we
should be able to get an independent estimate of the time-scales
involved in the SNIa production. However, the formation of carbon is
not yet well understood with the yields varying considerably between
different studies and objects (see e.g. \citealt{gustafsson},
\citealt{bf06}, and \citealt{carigi} for further discussions of
this). Also, the final interpretation of the abundance trend present
in the thick disk should be combined with results for the
halo. However, for the halo stars one need to rely on permitted carbon
lines. As these are subject to departures from local thermodynamic
modelling detailed modelling is required (see \citealt{fabbian06}).

\subsection{Manganese}

\begin{figure*}[t!]
\resizebox{\hsize}{!}{\includegraphics[angle=0]{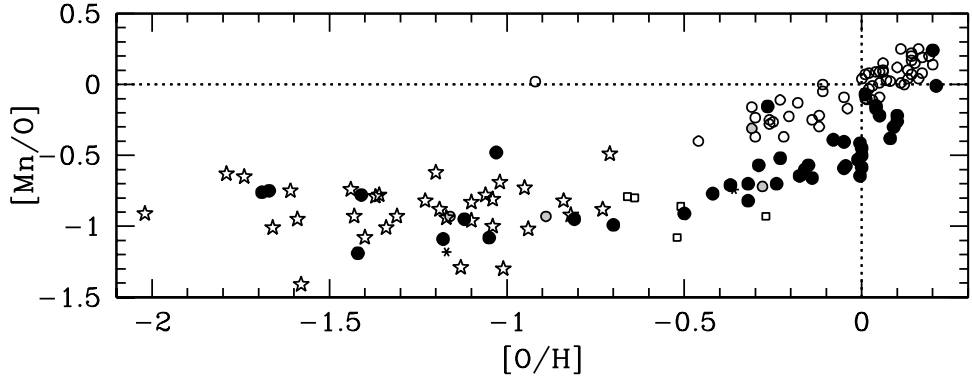}}
%\resizebox{\hsize}{!}{\includegraphics[ clip=true]{mno_mw.ps}}
\caption{\footnotesize
[Mn/O] vs [O/H] for the Galaxy. Here we distinguish the different kinematic
components instead of distinguishing the different studies. 
$\circ$ indicates stars with kinematics typical for the thin disk,
$\bullet$ indicate stars with kinematics typical of the thick disk, 
grey circles indicate 
transition objects between the thin and thick disk, 
$\star$ indicates stars with 
kinematics typical of the halo, and $\ast$ transition objects between the 
thick disk and the halo. 
$\square$ shows the five stars with oxygen abundances from \citet{melendez}.
No kinematic information is available for these stars.
Figure taken from \citet{feltzing06}
}
\label{mno}
\end{figure*}

\citet{feltzing06} have analysed four Mn\,{\sc i} lines in 95 dwarf
stars previously studied by \citet{bensby03} and \citet{bensby05}. The
stars were selected to have kinematics typical of the thick or the
thin disk. Using these two well defined and well studied stellar
samples we find that the abundance trends in the two samples differ
such that the stars with thin disk kinematics are enhanced in
manganese relative to stars with thick disk kinematics.

We also find that the previously reported ``step'' in the [Mn/Fe] vs
[Fe/H] trend for disk stars in the Galaxy (\citealt{mcwilliam03} and
\citealt{nissen00}) is in fact due to incomplete sampling of, in
particular, the more metal-rich part of the thick disk (compare the
discussion in \citealt{bensby03} of the kinematic selection of the
samples of stars studied in \citealt{nissen00} and
\citealt{chen00}). Thus there is no longer any need to invoke a large
spread in the age-metallicity relation for the thin disk to explain
the step.

Furthermore, when comparing the manganese abundances with iron
abundances the thick disk stars show a steadily increasing trend of
[Mn/Fe] vs. [Fe/H] whilst the stars with kinematics typical for the
thin disk show a flat trend up and until [Fe/H] = 0 and after that an
increasing trend.

In order to further study the origin of manganese we have combined the
new manganese abundances with oxygen abundances. Iron is made both in
SN\,II and in SN\,Ia. By using oxygen, which is only made in SN\,II,
as the reference element we simplify the interpretation of the
abundance data. For the stars in \citet{feltzing06} we took the oxygen
abundances from \citet{bensby04} and added manganese and oxygen data
from a number of other studies of (mainly) giant stars in the disks
and halo of the Galaxy. The full data set is shown in Fig.\,\ref{mno}.

For the halo and metal-poor thick disk, [O/H]$\leq -0.5$, the [Mn/O]
trend is flat.  This indicates that the production of Mn and O are
well balanced. Moreover, we know from the study of Bensby et
al.\,(2004) that the archetypal signature of SN\,Ias in the thick disk
do not occur until [O/H] = 0. Hence the up-going trend we see after
[O/H] $\simeq -0.5$ must be interpreted as being due to metallicity
dependent yields in SN\,II. The rising trend seen for the thin disk
sample could also be interpreted in this fashion. Although here we do
know that SN\,Ia contribute to the chemical enrichment and hence the
increase might also be due to these objects.  This is, essentially, in
agreement with the conclusions in \citet{mcwilliam03}.
 
\section{Impact of surveys}

The impact of future and on-going surveys on our understanding of the
abundance structure of the Galactic disk are two-fold; a) for surveys
that include enough spectral coverage gross abundance trends for the
different populations will be derived; b) provide catalogues from
which to select samples for high-resolution spectroscopy. With regards
to the abundance structure of the Galactic disk it is the latter
property that will be the most important.

One of the most exiting features of the big surveys is that we will be
able to identify suitable targets at large distances for differential
abundance studies with the local, on-going studies providing the
reference. With large, well defined samples of such far away dwarf
stars we will be able to grasp the abundance structure of the Galactic
disk in a way that is not possible at the moment being, as we are,
bound to the solar neighbourhood and the kinematically defined
samples.

\paragraph{What do we need from the surveys?}

What we need from the surveys in order to select targets for high
resolution spectroscopy is fairly straightforward and include:

\begin{enumerate}
\item photometry and/or low resolution
 spectroscopy 

\item distances, parallaxes, or evolutionary stage

\item radial velocities 

\item understanding of completeness
\end{enumerate}

The first item on the list will provide effective temperatures and (at
least for certain photometric systems as well as for low resolution
spectroscopy) estimates of the metallicities of the stars. The second
item is important for the derivation of surface gravity. Such
information is vital for the derivation of the final elemental
abundances. Often it is assumed that this property may be derived
directly from the measured equivalent widths by demanding that lines
arising from e.g. neutral and singly ionised iron give that same iron
abundance. However, this is not necessarily true for all evolutionary
stages and/or temperatures (\citealt{thevenin}). Hence an independent
estimate of this parameter is essential. The third item, together,
with distances, give kinematic information.

The last item on the list is perhaps the most important when it come
to the interpretation of the abundance trends. Some studies, notably
\citet{ap04}, have provided abundances for a volume complete
sample. However, these important studies also show the, current,
limitations in this approach: stars belonging to populations that have
a low density in the solar neighbourhood, e.g. the thick disk, have a
very small presence in the samples and hence any furthering of the
understanding of the, for example the thick disk, is limited. This is
also true for other small populations, such as the streams.

\paragraph{Everything directly from spectra?}
Would it not be desirable to directly get all the information discussed
above directly from the stellar spectra themselves? Indeed it would,
however, the problem with ionisational equilibrium discussed above
will naturally limit any such effort to narrow ranges of stellar
parameters for which ionisation equilibrium may be safely assumed (see
e.g. \citealt{ki03} and \citealt{bensby03} for two examples of such
parameter spaces).

\paragraph{Ages}

Determination of ages for individual stars is notoriously difficult
\citep{bjarne} and using stellar isochrones it is essentially only
stars just leaving the main-sequence that can be age dated with any
accuracy. Other methods, such as measurements of chromospheric
activity, may be used to date the younger dwarf stars. So far, all
studies that have looked at ages for stars typical of the thin and the
thick disk agree that the thick disk is on average older than the thin
disk (e.g. \citealt{fuhrmann04}, \citealt{bensby04b}, \citealt{gratton00}). The
debate about whether there is a gap in ages between these two population is
still an open question -- one that should be possible to resolve with
larger samples of stars for which we have good ages. Such studies will
also explore the age structure further away from the sun than has
hitherto been feasible (e.g. \citealt{nordstrom04}).

Also in studies of ages for different populations it is useful to take a
differential approach which means that the exact ages derived do not
matter as much and that the focus can be put on finding the
differences between populations, i.e. which came first and whether there are
over-laps in ages between the populations. Also here the surveys
should be able to provide us with samples reaching further away,
exploring the structure of the Galactic disk(s).

\section{Conclusions}

Today we know that, in the disk of the Galaxy, stars with distinct
kinematical signatures show distinctly different elemental abundance
trends. These findings have implications for models of galaxy
formation and evolution (both dynamical and chemical). However, so far
we have only been able to scratch on the surface of the abundance
structure of the Galaxy since we have been forced to mainly study the
solar neighbourhood. Surveys may provide us with well defined samples
of stars at large distances for which we may obtain stellar spectra
and hence compare their chemical composition to our nearby
neighbours. For success two things are essential. The first is that
the stars selected for these new studies have the same overall
properties as the stars we are studying in the local solar
neighbourhood. This will enable so called differential abundance
studies. Such studies have been proved to be very powerful also in
finding small differences between stellar populations as to first
order the errors made in in the abundance analysis cancel. The second
requirement for success is that these new samples are drawn from
stellar samples for which the sample characteristics (and in
particular the completeness) are well understood so that we know if we
are sampling the very rare parts of the Galaxy or major portions of
its stellar populations.

Ongoing studies are starting to explore stellar populations that are
not necessarily belonging to either the thick or the thin disk. For
example a study of the Hercules stream shows it to have elemental
abundance trends like the thick disk but a mean metallicity more like
that of the thin disk.

As recently stressed by \citet{fabbian06} an increased knowledge of
atomic data, 3-dimensional model atmospheres, and abundance
calculations taking departures from local thermodynamic equilibrium
into account are also necessary ingredients if we want to understand
the formation and evolution of the stellar populations in the
Galaxy. Hence, large surveys combined with detailed studies of
smaller, representative samples and improved techniques of analysis
are all equally essential ingredients for future progress.
 
\begin{acknowledgements}
SF is a Royal Swedish Academy of Sciences Research Fellow supported
by a grant from the Knut and Alice Wallenberg Foundation. The IAU is
thanked for a grant that contributed towards the cost of the
authour's participation in the Joint Discussion 13 at the General
Assembly held in Prag 2006.
\end{acknowledgements}

\bibliographystyle{aa}

\end{document}